# Liquid Crystal Elastomer – Liquid Metal Composite: Ultrafast, Untethered, and Programmable Actuation by Induction Heating

*Victor Maurin, Yilong Chang, Qiji Ze, Sophie Leanza, Ruike Renee Zhao\**

Department of Mechanical Engineering
Stanford University
Stanford, CA 94305, USA
E-mail: rrzhao@stanford.edu

**Abstract:**

Liquid crystal elastomers (LCEs) are a stimuli-responsive material which has been intensively studied for applications including artificial muscles, shape morphing structures, and soft robotics, due to its capability of large, programmable, and fully reversible strains. To fully take advantage of LCEs, rapid, untethered, and programmable actuation methods are highly desirable. Here, we report a liquid crystal elastomer – liquid metal (LCE – LM) composite, which enables ultrafast actuations and high heating programmability by eddy current induction heating. The composite consists of LM sandwiched between two 3D-printed LCE layers via direct ink writing (DIW). When subject to a high-frequency alternating magnetic field, the composite can be actuated in milli-seconds. By moving the magnetic field, the eddy current can be spatially controlled for selective actuation. Additionally, sequential heating is achievable by programming the LM thickness distribution in a specimen. With these capabilities, the LCE – LM composite is further exploited for multimodal deformation of a pop-up structure, on ground omnidirectional robotic motion, in water targeted object manipulation, and crawling.

**Introduction**

Stimuli-responsive materials capable of altering their shape and physical properties in response to external stimulations, such as magnetic fields, light, or heat, have enabled a range of novel soft robotic capabilities, including compliant deformation,[1–3] bioinspired movement,[4–6] and adaptive motion in extreme and unpredictable environments.[7,8] Examples of stimuli-responsive materials include shape memory polymers,[9–12] hydrogels,[13–15] soft magnetic polymers,[16–18] and liquid crystal elastomers (LCEs).[19–21] LCEs have been largely explored to develop artificial muscles,[22,23] shape morphing structures,[24–26] and soft robots[27,28] due to its capability of generating large, programmable, and fully reversible strains up to 50-60% under temperature change.[29] When an LCE is heated above its transition temperature, a nematic-to-isotropic phase transition is triggered inducing a microstructural rearrangement of its mesogens from an aligned state to a disorganized state, producing a macroscopic contraction along the mesogens alignment direction.[30] With this effective thermo-mechanical response, LCEs possess advantages of generating reversible and large shape changes and high energy density deformations.[31]

To drive the thermal actuation of LCEs, joule heating,[32–34] photothermal,[35,36] infrared (IR) heating,[37] heat guns,[38] hot plates and hot water baths,[26,39] have been used to increase the temperature. Among these methods, joule heating has been widely utilized to program the heating distribution in LCE structures for desired shape morphing by embedding thermal resistive components such as heating wires and pads at the prescribed locations in the LCE. Yet, this method is tethered, and the heating components are usually rigid, which constrains the deformation of the LCE and potentially causes delamination.[32–34] For methods using heat guns, hot plates, and hot water baths, they increase the temperature of the whole specimen at once allowing for only a single deformation mode.[26,38,39] In comparison, IR and photothermal heating can achieve more localized energy transmission to a specific region of the LCE for selective actuation. However, they still possess limitations when considering actuation environments that are inaccessible by IR or light.[40]

Alternatively, induction heating is a promising untethered strategy for rapid temperature increase of materials. This method uses high frequency electromagnetic fields and can be potentially integrated into LCEs for untethered and programmable heating. In general, there are two induction heating mechanisms, namely hysteresis loss and eddy current. Hysteresis loss converts the energy loss from the magnetization and demagnetization cycle of ferromagnetic materials to produce heat. Heating soft materials through hysteresis loss is usually implemented through embedding magnetic particles, such as iron oxide particles ($Fe_3O_4$), in the soft polymer matrix .[9,41–43] However, such method has obvious drawbacks due to its limited heating efficiency, the significant increase in polymer stiffness, and the destruction of the polymer chain alignment during the fabrication, which hinder LCE deformation.[44,45] On the other hand, eddy current generates heat through the electrical current induced in a conductive material under a changing magnetic field. Considering liquid metal (LM)[46] as the conductive material for induction heating of thermal-responsive soft materials such as LCEs, it offers tremendous advantages as it provides ultrafast heating for rapid and large actuation, in the meantime does not constrain the deformation of the soft system due to LM's negligible stiffness.

In this work, we report a novel liquid crystal elastomer – liquid metal (LCE – LM) composite which allows for ultrafast, untethered, and highly programmable actuation through eddy current induction heating. The composite consists of LM sandwiched between two 3D-printed LCE layers. The LCEs are printed via direct ink writing (DIW) and thus their actuated configurations can be controlled through the printing pathway.[38,47,48] The LM is sprayed with a mask, which allows for programmable patterns.[49–51] The eddy current heating efficiency can be regulated by the number of LM layers sprayed. When subject to a high frequency alternating magnetic field, the composite can be actuated in milli-seconds. By moving the magnetic field, the eddy current can be spatially controlled to enable selective actuation. Additionally, sequential heating is achievable by programming the LM thickness distribution. The LM – LCE composite, together with the 3D printing capability, is further exploited for multimodal deformations of a pop-up structure, on ground omnidirectional robotic motions, in-water targeted object manipulation, and in-water crawling. This work presents a novel way for ultrafast and untethered actuation of LCEs for applications where remote control is highly desirable.

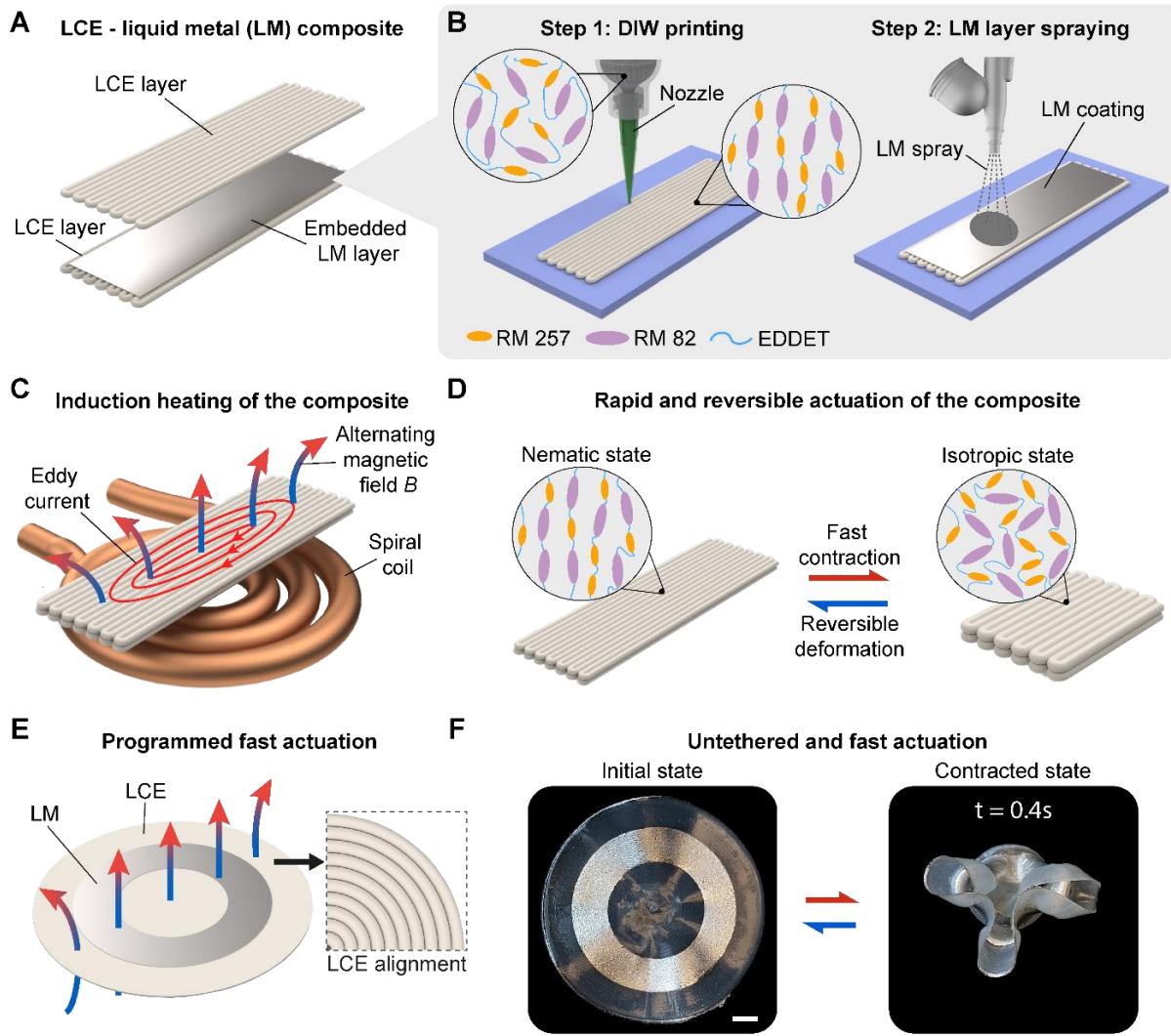

**Figure 1.** LCE – LM composite manufacturing, induction heating mechanism, and programmed fast actuation. (A) Exploded view of the LCE – LM composite consisting of LM sandwiched between two LCE layers. (B) DIW printing and LM spraying of the printed LCE layer of the composite. The mesogens are aligned along the printing path. (C) Schematic of induction heating of the composite, where a high frequency alternating magnetic field induces eddy current in the LM. (D) Schematic of rapid contraction and reversible deformation. (E) Induction heating schematic of an LCE disk with an LM ring sprayed on top. The alignment of LCE mesogens is in the circumferential direction. (F) Experimental images of rapid actuation of the disk through induction heating. Scale bar: 5 mm.

## 2. Results and Discussions

### 2.1. LCE-LM composite design and fabrication

The manufactured composite (**Figure 1A**) is composed of LM sandwiched between two 3D-printed LCE layers, which are printed via DIW. The printed LCE has a measured shear modulus of 493 kPa (see Figure S1 for more details on the mechanical characterizations). As illustrated in **Figure 1B**, to allow for programmable shape morphing, the LCE mesogens are aligned through the shear stress generated during the extrusion of the LCE fibers from the print head. The LCE deformation can be designed by programming the DIW print path. The printed layers are cured with UV light (see experimental section and Figure S2 for more details on the composite manufacturing). The LM is then sprayed on top of the printed LCE with a mask, which allows for different LM patterns (**Figure 1B**). Finally, another LCE layer is printed and placed on top to cover the LM, and the whole structure is cured under UV light. When a high frequency alternating magnetic field is applied perpendicularly to the LM sheet, ohmic loss is generated through eddy current (**Figure 1C**). The temperature can increase to over 100 °C in milliseconds, during which the LCE transitions from its nematic state to isotropic state for ultrafast, large, and reversible contraction (**Figure 1D**). **Figure 1E** illustrates an LCE – LM composite that is made of a circular LCE sheet with circumferentially aligned mesogens and an LM ring patterned on top. The LCE – LM sheet demonstrates exceptionally fast actuation in 0.4s upon induction heating (**Figure 1F**, Movie S1). Utilizing this untethered, programmable, and fast heating strategy, we design LCE – LM composites to enable sequential and selective actuation for functions including multimodal deformation, on-ground, and in-water robotic motion.

### 2.2. Induction heating characterizations of LM

The heat generated by induction heating of LM is determined by the LM pattern geometry, thickness, as well as the magnetic field intensity and frequency. In this section, the induction heating efficiency of LM is first characterized by heating LM squares of the same size (15 mm x 15 mm) but with different thicknesses. Note that the thickness of LM is controlled by the number of coatings sprayed, denominated as "layers", with a single layer being approximately 10 µm thick (See SI for more details on LM thickness characterizations). As depicted in **Figure 2A**, an LM square with five layers (~50 µm thick) is first heated under a nearly uniform magnetic field with amplitude $B$ = 45 mT and frequency $f$ = 51 kHz (See Figure S3 for more information on the coil). The temperature distribution recorded by IR imaging agrees with the simulated ohmic loss distribution, indicating higher temperatures at the edges due to the skin effect generated by the high frequency alternating magnetic field[52,53] (see Movie S2 for more information on the heating process, and SI for more details on the ohmic loss simulation). As shown in **Figure 2B**, the ohmic loss increases nearly linearly with the thickness under the same alternating magnetic field. In **Figure 2C,** it can be observed that the heating speed increases with the numbers of LM layers. With 8-layer LM (~ 80 µm thick), the temperature increases to above 100 °C in less than one second. With 30-layer LM, (~ 300 µm thick), it reaches 500 °C in five seconds, demonstrating extremely fast heating. For LCE actuation, 100 °C can provide 30% strain. Therefore, for a composite with more than eight layers of LM, the LCE is able to reach 30% contraction in less than one second (See Figure S1 for the strain vs temperature curve of the LCE).

Induction heating makes on-demand selective actuation easily achievable by changing the location of the electromagnetic coil. As illustrated in **Figure 2D**, a smaller spiral coil (see Figure S4 for details on the coil) is used to generate a concentrated alternating magnetic field with amplitude $B$ = 144 mT, and frequency $f$ = 46 kHz. By placing the coil at three distinct positions, localized heating occurs in these targeted regions, which is demonstrated by heating different locations of the two vertically positioned 7-layer LM squares (See Movie S2 for more information on the induction heating of LM squares). The temperature distribution recorded by IR imaging is accurately predicted by the simulated ohmic loss. When the coil is placed at position 1, under the top square, concentrated heating occurs only for the LM of the top square, and no heating is observed on the bottom LM square. When the coil is moved to position 2, in between the two LM squares, the heating is seen at the edges of the two LM squares. Finally, at position 3, under the bottom square, heating of the LM is only observed for the bottom LM square, whereas no heating is observed for the top LM square.

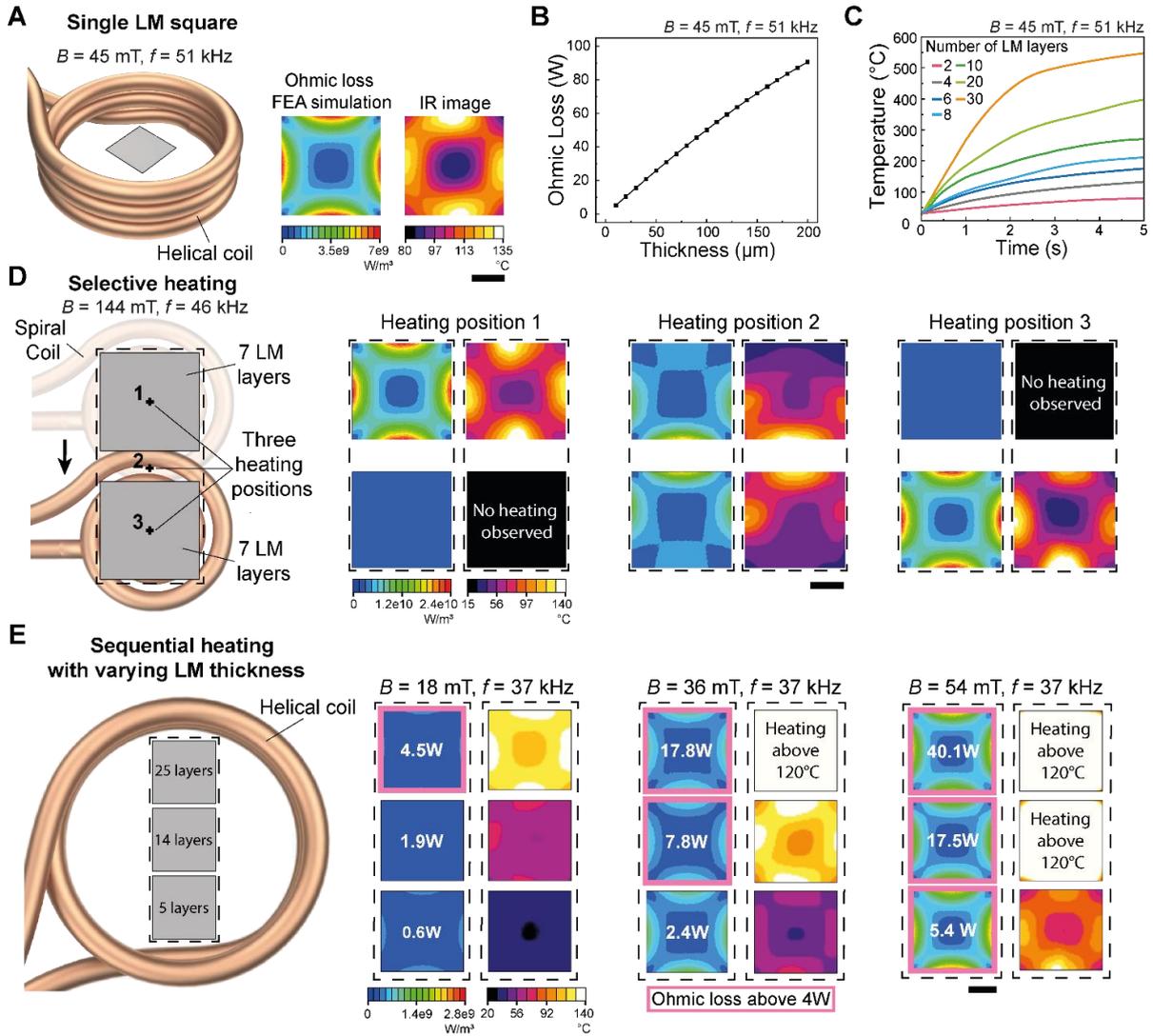

**Figure 2.** Characterizations of induction heating of LM squares (15 mm × 15mm) for selective and sequential heating. (A) Induction heating of a 5-layer LM square and its FEA-predicted ohmic loss distribution and measured temperature distribution under $B$ = 45 mT and $f$ = 51 kHz. (B) Ohmic loss of the LM square vs thickness/number of LM layers. (C) Measured temperature vs time for the LM square with varied thickness/number of LM layers. (D) Ohmic loss and measured temperature distribution during selective heating of two 7-layer LM squares when a coil is placed at three distinct positions labeled 1, 2 and 3 under $B$ = 144 mT and $f$ = 46 kHz. (E) Ohmic loss and measured temperature distribution during sequential heating of three LM squares with varied numbers of layers (25, 14, 5 layers) under increasing magnetic field intensity at $f$ = 37 kHz. Scale bars: 5 mm.

Additionally, since the induction heating speed is proportional to the number of LM layers, sequential actuation is feasible by gradually increasing the magnetic field to regions with different LM layer numbers. To demonstrate this, three LM squares with varied thicknesses are sequentially heated above 120 °C inside a helical coil (Figure S5 for more details on the coil) at different magnetic field strengths at $f$ = 37 kHz, as illustrated in **Figure 2E**. Here, 120 °C is the temperature for LCE to reach 40% of actuation strain (See Figure S1 for the strain vs. temperature curve of the LCE). The goal is to reach 120 °C for the top LM square under $B_1$ = 18 mT, the top two squares under $B_2$ = 36 mT, and finally all three LM squares, under $B_3$ = 54 mT. To achieve this, the number of LM layers on each square is calculated by the specific heat equation with the critical ohmic loss set as 4W, above which a square would be able to reach 120 °C in less than 2 s, for fast actuation (see the supplementary information for more details on the calculations of the critical ohmic loss). Based on the calculation, the three squares (positioned vertically in **Figure 2E**, from top to bottom) should consist of 25 (~250 µm thick), 14 (~140 µm thick), and 5 (~50 µm thick) layers. It can be seen that, when the magnetic field increases from 18 mT, to 36 mT, and finally to 54 mT, the three squares are sequentially heated above 120 °C in three steps, starting from the top square with the highest number of LM layers (top square), to the top two squares, and lastly to all three squares (see Movie S2 for more details on the heating process). It can also be seen that only the squares with an ohmic loss higher than 4W achieves heating above 120 °C at each step. The demonstrated selective heating and sequential heating are further utilized for programmable deformation and motions in the following sections.

## 2.3. In-water motion and object manipulation by programmable heating

The untethered induction heating of LM allows for fast and reversible actuations of the LCE in room temperature water, especially because of the fast cooling provided by the high thermal conductivity of water compared to air. The easy repositioning of the magnetic field by moving the coil further enables the spatially programmable eddy current for controllable LCE motion and function in aquatic environment, which has never been achieved before. To demonstrate the programmable in-water actuation of the LCE – LM composite, we manufactured an LCE – LM bilayer, which is composed of six LM layers (~60 µm thick) sandwiched between two LCE stripes with mesogens aligned longitudinally by DIW, as shown in **Figure 3A**. The top surface of the stripe is then bonded with a layer of a thermally inactive material (shear modulus 870 kPa, see Figure S1 for more details on the mechanical characterizations), which causes the sample to bend when the LCE layers are actuated (More details on the bilayer manufacturing can be found in the experimental method section). Even though the LM pattern is a long rectangle, it is still very effective to actuate a selected LCE region by applying a localized magnetic field. As demonstrated in **Figure 3B**, the ohmic loss (FEA predicted) and the resulted high temperature zone (IR measured) travels along the LCE stripe with the moving coil, while the coil generates an alternating magnetic field with maximum magnitude $B = 144$ mT, $f = 46$ kHz (see Movie S3 for more details on the heating, see Figure S4 for more details on the coil). A wave propagation motion of the bilayer stripe is observed when the coil is moving along the stripe, as illustrated in **Figure 3C**. The waves front moves simultaneously with the coil, demonstrating efficient heating even in cold water. The fast cooling of water also allows for rapid motions (see Movie S3 for more details on the wave propagation). As shown in **Figure 3D**, this motion is also well captured by FEA simulations. To model the bilayer bending, we imposed a nonuniform temperature distribution to replicate the temperature profile observed experimentally (See Figure S8 and FEA section in the supplementary information for more details). The resulting bending deformation and wave propagation closely matches the continuous motion observed in experiments (see Movie S3 for more details on the FEA deformation).

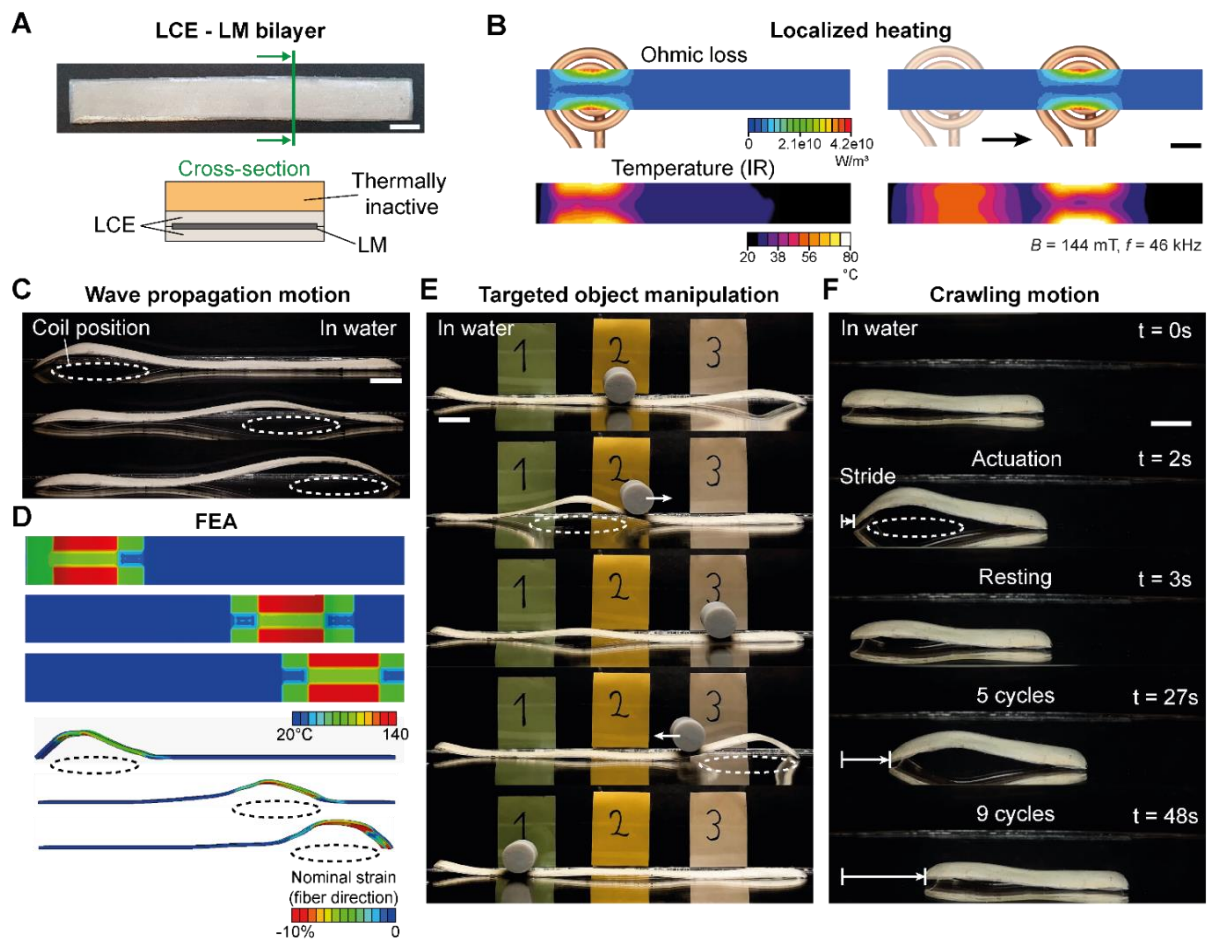

**Figure 3.** In-water actuation of LCE – LM bilayer stripes for wave propagation motion, object manipulation, and crawling. (A) Manufactured sample and its material composition shown by the cross-section schematics. The sample consists of two longitudinally aligned mesogens LCE layers with 6-layer LM sandwiched in between. A thermally inactive material layer is bonded on top of the LCE – LM composite. Scale bar: 10 mm. (B) Simulated ohmic loss distribution and measured temperature distribution of the LM locally heated by a spiral coil with $B$ = 144 mT and $f$ = 46 kHz. Scale bar: 10mm. (C) Wave propagation motion achieved by moving the coil underneath the sample in water. Scale bar: 10 mm. (D) Temperature distribution in thermo-structural FEA and simulated wave propagation motion of the sample. (E) In-water manipulation of a cylinder through localized heating of the sample. Scale bar: 10 mm. (F) In-water crawling motion through repeated heating and cooling of the sample. Scale bar: 7 mm.

This move propagation motion can be further utilized to precisely manipulate an object in water. As demonstrated in **Figure 3E**, the wave motion induced by the localized actuation of the LCE – LM stripe accurately pushes a cylinder from one target position to another (see Movie S3 for more details on the targeted object manipulation). Finally, a crawling motion is achieved in water through the cyclic actuation of a shorter stripe. As depicted in **Figure 3F**, for each actuation cycle of the crawler, the heat is localized on the left end of the sample, causing only the left part to bend and lift from the ground. The high friction of the undeformed right part of the crawler results in the bending-induced contraction of the sample towards the right. The coil is then moved to the right, and the recovering motion of the left part pushes the whole body to translate to the right. After nine cycles of reversible bending deformation, the sample moves by approximately half of its bodylength in water in less than one minute (see Movie S3 for more details on the crawling motion). In summary, spatially programmable eddy current induction heating coupled with the fast cooling of water enables agile motions of the LCE in water, leading to controllable aquatic object manipulation and crawling.

**2.4. Sequential actuation of a pop-up structure**

The intensity of eddy current can also be spatially programmed for sequential actuation of the LCE – LM composite by prescribing the LM thicknesses distribution in the specimen. In this section, a sequential popping-up motion is demonstrated by designing regions with different numbers of LM layers. **Figure 4A** shows the manufactured LCE – LM circular disk, with mesogens aligned circumferentially by DIW. To show the sequential actuation, the circular disk is divided into two regions by the LM patterns: an inner LM circle and an outer segmented LM ring. Note that the ohmic loss of the LM is determined by both its geometry and thickness. To precisely program the ohmic loss in these two LM regions, the area of the inner circle is designed to be the same as one segment of the outer ring, so that they induce the same ohmic loss when having the same LM thickness. Thereafter, the ohmic loss of the two regions can be accurately controlled by programming their numbers of LM layers separately. As shown in **Figure 4B,** there are twenty layers of LM (~200 µm thick) for the inner circle and six layers of LM (~60 µm thick) for the segmented LM ring. The LM is sandwiched between a thinner bottom and a thicker top circular LCE layer. The thickness difference of the two LCE layers enables the bending in a prescribed direction, as the thinner LCE layer will contract more when the structure is actuated. As illustrated in **Figure 4C**, the pop-up structure is heated by a spiral coil (See Figure S6 for more details on the coil), with frequency $f$ = 60 kHz. The structure is placed on a substrate above the coil. The inner LM circle and the outer LM ring patterns are highlighted by green and purple dashed lines, respectively, along with two reference points, point A (red) and point B (blue), which are used to measure the deformation of the structure throughout its actuation.

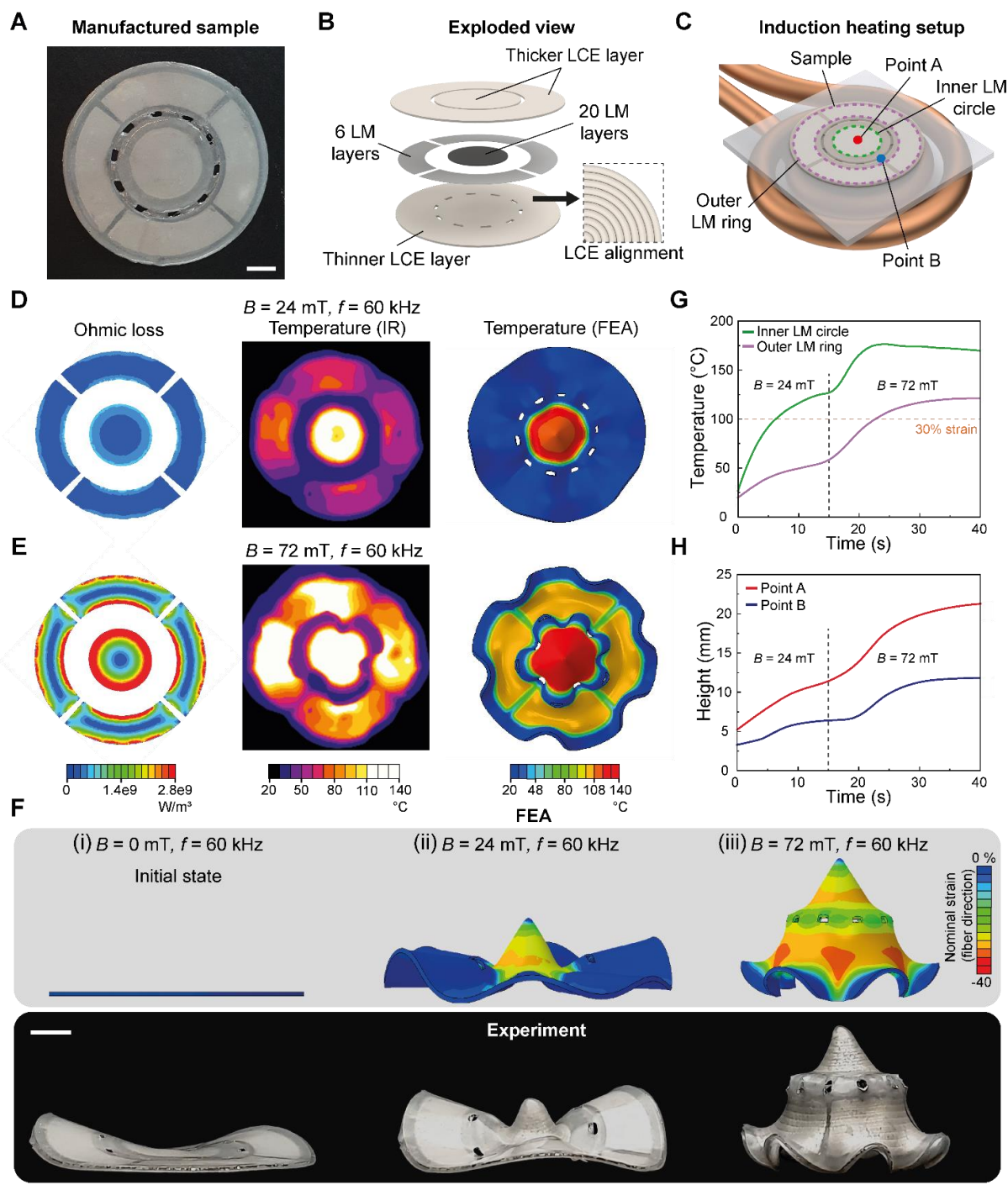

**Figure 4.** Pop-up LCE – LM structure and its sequential actuation. (A) Manufactured sample. Scale bar: 5 mm. (B) Exploded view of the sample, with a 20-layer LM for the inner LCE circle and a 6-layer segmented LM ring for the outer LCE ring. The LCE layers have circumferential mesogens alignment. (C) Induction heating setup. The sample is placed on a substrate laying on a spiral coil. Point A, point B, the inner LM circle and the outer LM ring are represented on the sample. (D) Simulated ohmic loss distribution FEA, measured temperature distribution, and temperature distribution in thermo-structural FEA for $B = 24$ mT and $f = 60$ kHz, and for (E) $B = 72$ mT and $f = 60$ kHz. (F) Sequential actuation (FEA and experiment) under the three magnetic fields (i) $B = 0$ mT, (ii) $B = 24$ mT, and (iii) $B = 72$ mT. Scale bar: 5 mm. (G) Temperature vs time for the inner LM circle (green) and the outer LM ring (purple). The vertical black dashed line denotes the time at which the magnetic field is increased from 24 mT to 72 mT. The horizontal orange dashed line represents 100 °C, which generates 30% LCE strain. (H) Height at points A (red) and B (blue) vs time during actuation. The vertical black dashed line denotes the time at which the magnetic field is increased from 24 mT to 72 mT.

To achieve sequential heating, a magnetic field is first gradually ramped up to $B = 24$ mT and $f = 60$ kHz, under which the ohmic loss distribution induces efficient heating actuating only the inner LM circle region of the LCE structure. This selective actuation is explained by the ohmic loss and IR measurement of the temperature shown in **Figure 4D**. As depicted in this figure, the ohmic loss is higher at the center of the sample, and it exceeds the critical value calculated as 1.4 W, corresponding to a temperature increase of the LCE above 80 °C (15% strain) to generate visible deformations in 2 s (See SI for more details on the specific heat calculations, and Figure S1 for the temperature vs. strain curve of the LCE). However, the outer LM ring fails to reach the critical ohmic loss and temperature increase to induce sufficient strain for visible shape change. As shown in **Figure 4E**, upon increasing the magnetic field magnitude to $B = 72$ mT under the same frequency, the ohmic loss is sufficiently high in both LM regions to increase the temperature above 120 °C for large shape change. Through this two-step actuation, sequential popping-up deformations are obtained with this circular LCE – LM composite with programmed LM thickness distribution. To predict the sequential deformation, we adopt the measured temperature observed in the sample at the two magnetic field amplitudes and input them in the thermo-structural FEA. The simulated temperature distribution shown in **Figure 4D** and **Figure 4E** is resulted from the heat transfer process. The comparison between the simulated sequential pop-up behavior and the actual actuation is shown in **Figure 4F**, demonstrating excellent agreement (see Movie S4 for more details for the FEA and experiment). The initial state, first actuated state, and the final actuated state of the structure is shown in **Figure 4F**(i), (ii), and (iii), at $B = 0$, 24, and 72 mT, respectively. For details on the thermo-structural simulation, see supplemental information section in SI. It is also worth noting that the non-flat initial state of the manufactured structure is attributed to the residual stress during sample fabrication. This initial state does not affect the sequential actuation, due to the large contraction of LCEs. This also confirms the robustness of the current design and actuation approach.

To characterize the temperature evolution in the different regions of the structure, the temperature at the center of the inner LM circle and the average temperature of the centers of the outer LM rings are recorded during heating, as illustrated by the green and purple curves in **Figure 4G**, respectively. It can be seen that within the first 15 s, the structure is actuated only in the inner LM circle region, under $B = 24$ mT. The temperature achieved in this region surpasses 100 °C, corresponding to 30% strain of the LCE, (denoted by the orange dash line) and thus enables almost full contraction along the LCE fiber direction, enabling a significant actuation of the inner section of the structure, which is demonstrated by the deformation side view in **Figure 4F**(ii) (see Figure S1 for the temperature vs. strain of the LCE). On the other hand, the outer LM ring temperature plateaus around 50°C, which is insufficient to trigger large LCE deformation. However, when the magnetic field is increased to $B = 72$ mT at 15 s, both the inner LM circle and the outer LM ring surpass 100 °C, allowing the whole structure to be actuated, as demonstrated by the deformation side view in **Figure 4F**(iii). The deformation of the sample at different positions is then quantitatively measured and presented in **Figure 4H**, in which the elevation of point A and point B is recorded during pop-up of the structure, as illustrated by the red and blue curves respectively. For $B = 24$ mT, the inner LM circle region of the structure, represented by point A, elevates at a higher rate compared to the outer LM ring region, represented by point B. However, when $B$ is increased to 72 mT, the heights of both A and B increase significantly (see Movie S4 for more information on the pop-up deformation). The sequential popping-up of the LCE – LM structure shows the easy-programming of multimodal deformation by designing the LM heating efficiency through its thickness distribution.

### 2.5. Robotic sea turtle with omnidirectional locomotion

In this section, we design a soft robot capable of controllable omnidirectional crawling to mimic motions of a sea turtle through the actuation of LCE – LM "fins". The turtle robot consists of a 3D-printed rigid body driven by two soft LCE – LM fins. As shown in **Figure 5A**, both triangulated fins are fabricated with twenty LM layers (~200 μm thick) sandwiched between the thick top and the thin bottom LCE sheets with mesogens aligned circularly by DIW (see experimental methods for more details on the fabrication of the fin). The turtle's motions are driven by the actuation of its fins, which undergo a series of deformation steps as illustrated in **Figure 5B**. Upon induction heating, the LCE sheets contract around the center of the circular alignment. The resulting deformation behaves like joints centered at this same point. At the first actuation step, the temperature distribution in the LCE sheets is asymmetric because of their thickness difference, causing the bending of the fins, which generates a pushing motion. As the heating increases, the fins reach their second deformation step, at which they are now fully actuated. It results in the fins deformation to occur in a plane almost parallel to the substrate. When cooling down, the fins do not follow the same deformation back as for the heating, as they do not bend but simply rotate their tips back around the joint center to their initial positions (**Figure 5B**, see Movie S5 for more information on the fin deformation). Due to their asymmetric deformation during heating and cooling, the fins are able to generate translational motion to move the sea turtle.

To demonstrate the turtle motion, the two triangular fins are glued to the 3D printed rigid turtle body (**Figure 5D**, see experimental methods for more details on the sea turtle manufacturing). A turning motion can be achieved by selectively actuating one fin at a time (**Figure 5E**), while a simultaneous actuation of both fins can lead to a forward motion (**Figure 5F**). As detailed in **Figure 5E**, we demonstrate the turtle turning motion by repeatedly actuating the left fin, and then the right fin, with a local magnetic field of amplitude $B = 108$ mT and frequency $f = 37$ kHz (see Figure S7 for more details on the coil). This turning motion is very well predicted by the thermo-structural FEA, in which a temperature of 140 °C is imposed in the fins (see FEA section in the SI for more details). The forward motion of the turtle is achieved by actuating both fins using a larger coil that provides a magnetic field spatially covering the two fins, with $B = 108$ mT, and $f = 37$ kHz (see Figure S7 for more details on the coil). The turtle is capable of walking along a straight-line, as shown in **figure 5F**, with good agreement between the FEA and experiment. Finally, As shown in **Figure 5G**, an "S"-shaped crawling path is realized by coupled turning and straight walking motions, demonstrating the turtle's ability to perform omnidirectional motions.

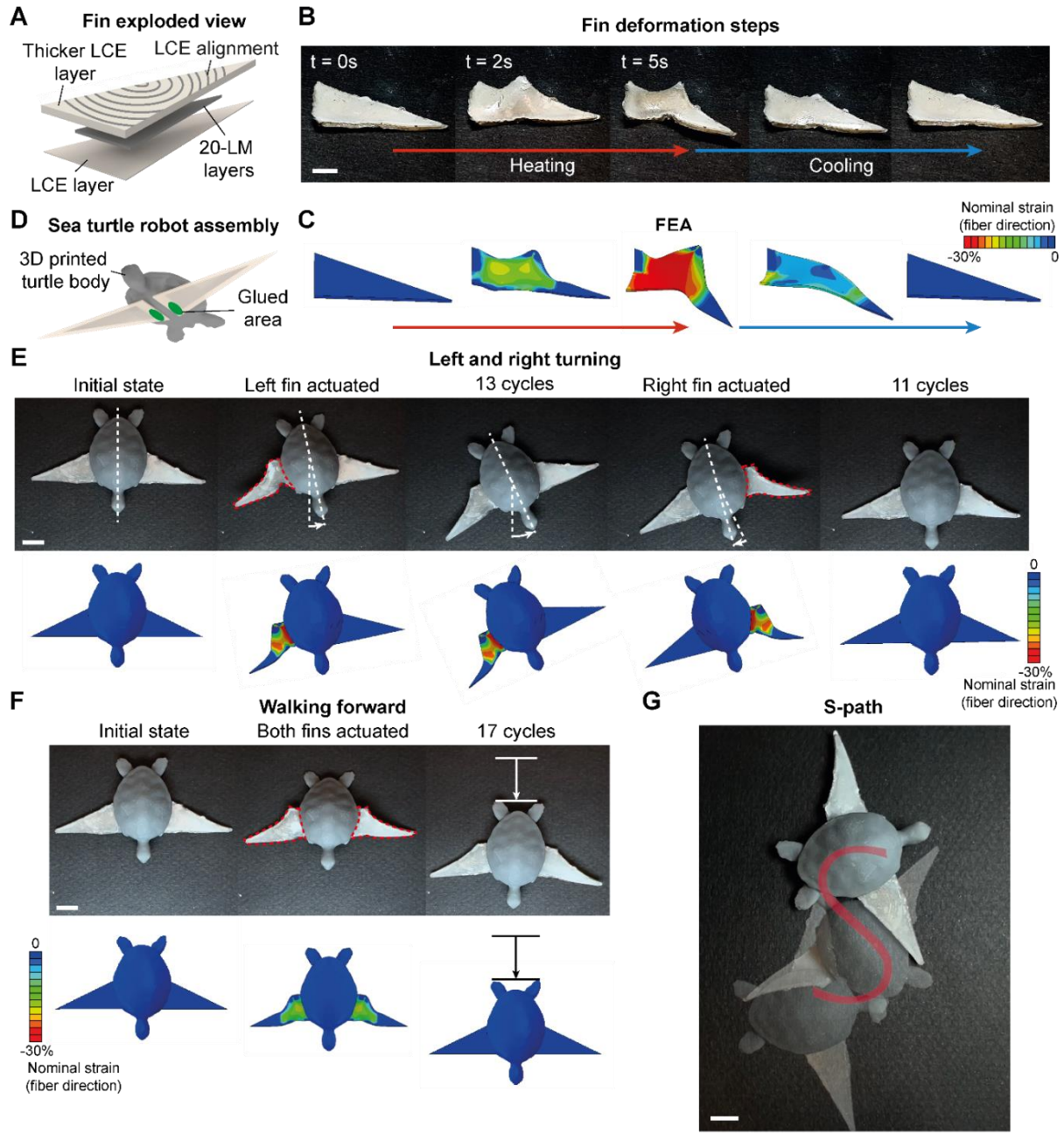

**Figure 5.** Actuation and omnidirectional motion of the LCE – LM robotic sea turtle. (A) Exploded view of a LCE – LM fin. The fin consists of 20-layer LM sandwiched between one thick and one thin LCE layer. The mesogens alignment of LCE is circular. (B) Fin actuation process. Scale bar: 5 mm. (C) Fin deformation predicted by thermo-structural FEA. (D) Schematic of the sea turtle assembly. The fins are glued onto the turtle body at the green zones. (E) Sea turtle turning motion (Scale bar: 7 mm) and thermo-structural FEA prediction. (F) Experiment (top, scale bar: 7 mm) and thermo-structural FEA (bottom) of the sea turtle straight-line walking motion. (G) Combined walking and turning motion of the sea turtle to follow an "S"-shaped path. Scale bar: 7 mm.

## 3. Conclusion

In summary, the novel LCE – LM composite reported in this work is capable of ultrafast, untethered and highly programmable actuations, enabled by eddy current induction heating. By moving the magnetic field and varying the LM thickness, the eddy current can be locally tuned, allowing for selective and sequential heating of the composite. The use of DIW to print the layers LCE also allows to easily modify the mesogens alignment direction for complex shape morphing. With those advantages, the LCE – LM composite has been successfully implemented for in water agile motions, such as targeted object manipulations and crawling, pop-up deformations, and omnidirectional robotic motions. Combined with the simulation tools that have been able to accurately predict the composite deformations, we believe that this work will permit the extension of LCE capabilities.

## 4. Experimental Section/Methods

*LCE – LM composite manufacturing:* To manufacture the LCE ink, the mesogens (1,4-Bis-[4-(3-acryloyloxypropyloxy)benzoyloxy]-2-methylbenzene) (RM 82, BOC Sciences, USA) and (2-Methyl-1,4-phenylene bis(4-((6-(acryloyloxy)hexyl)oxy)benzoate)) (RM 257, BOC Sciences, USA) are mixed together in a weight ratio of 3:1. Then, the spacer (2,2-(Ethylenedioxy)diethanethiol) (EDDET, Sigma Aldrich, USA) and (2,6-Di-tert-butyl-4-methylphenol) (BHT, Sigma Aldrich, USA) are introduced with 25 wt% and 2 wt%, respectively, compared to the total weight of mesogens. After melting all the components at 80 °C for 1 h, the catalyst triethylamine (TEA, Sigma Aldrich, USA) and the photoinitiator Irgacure 819 (Sigma Aldrich, USA) are added to the mixture with 1 wt% and 2 wt%, respectively, compared to the total weight of mesogens. The mixture is then stirred for 3 min with a magnetic stir bar at 80 °C for homogenization and placed in an oven at 80 °C during 25 min for oligomerization. Next, the mixture is transferred to a 10 mL syringe barrel (Nordson EFD, USA), and heated again in a vacuum oven at 80 °C for 20 min. The ink is finally defoamed in a planetary mixer (AR-100, Thinky, USA) at 2200 rpm for 3 min to remove trapped air.

For the printing of LCE, the syringe barrel filled with LCE ink is mounted to a customized gantry 3D printer (Aerotech, USA). The air pressure to each barrel is individually powered by a high-precision dispenser (7012590, Ultimus V, Nordson EFD, USA). The print path is designed using Solidworks (Dassault Systèmes, France) and converted to G-code from CADFusion (Aerotech, USA). The material is printed at room temperature at a speed of 15 mm/s and pressure of 550 kPa to align the LCE mesogens along the print path for large strain. The distance from the syringe to the substrate is approximately 0.2 mm, and the distance from two neighboring printed lines is 0.35 mm. To fix its printed filaments, the LCE is cured by UV light (385 nm) for several minutes after printing. The LCE shear modulus obtained after curing is 493 kPa (see Figure S1 for more details on the mechanical properties of LCE).

To create the LCE – LM composite, the LM (Galinstan, Rotometals, USA) pattern is first sprayed onto a cured LCE layer with a laser-cut paper mask, using an airbrush paint sprayer (Master Airbrush, USA) with a pressure of 1.5 bar. To apply a specific thickness of LM, we normalize the spraying procedure by defining "LM layers", with one layer corresponding to a coating sprayed at approximately 5 cm above the sample during 1 s. We then applied several layers depending on the desired thickness. Once coated, the sprayed LCE layer is then recovered by another LCE layer, and the system is finally cured by UV light (385 nm) for several minutes to firmly fix the two LCE layers together (See Figure S2 for more details on the composite manufacturing).

*LCE – LM bilayer manufacturing:* To manufacture the LCE – LM bilayers demonstrated in **Figure 3**, a LCE – LM composite is made following the same method described in *LCE – LM composite manufacturing*. The LCE layers printed with DIW have longitudinally aligned mesogens. The LCE layers used in **Figure 3A-E** are rectangles of 120 mm by 15 mm by 0.2 mm size, while the shorter layers used in **Figure 3F** are rectangles of 35 mm by 15 mm by 0.2 mm size. For each sample, the 6-layer (~60 µm thick) LM pattern is a rectangle offset by 1.5 mm from every edge of the LCE layers. To obtain the bending deformation, the composite is recovered by a thermally inactive photocurable layer capable to bond with LCE (Elastic 50A Resin, Formlabs, USA) with shear modulus of 870 kPa (see SI for more information on the mechanical characterization of the thermally inactive material), by spin-coating at 600 rpm for 6 s. To fix the whole structure together, UV light (385 nm) is shined for several minutes onto the bilayer.

*Sea turtle manufacturing:* To create the sea turtle fins, a LCE – LM composite is made following the same method described in *LCE – LM composite manufacturing*. The LCE layers are right-angled triangles of size 15 mm by 25 mm by 29 mm, and thickness 0.6 mm for the top layer, and 0.2 mm for the bottom layer. The LCE mesogens alignment is circular for both the bottom and the top LCE layers, and the alignment center is positioned on the triangle hypotenuse, at 8 mm from the shorter edge of the triangle. The LM pattern is also a right-angled triangle offset by 1.5 mm from every edge of the LCE layers and consists of twenty LM layers (~200 µm thick). During the manufacturing of the composite, a small amount of uncured LCE is applied around the LM pattern of the coated LCE layer to improve the bonding of the two LCE layers together. To fix the structure of the fins, UV light (385 nm) is shined for several minutes. The two fins are then fixed onto a 3D-printed sea turtle body (Grey Resin, Formlabs, USA) with a thermally inactive photocurable material capable to bond to LCE and to the turtle body (Elastic 50A Resin, Formlabs, USA), applied on the regions highlighted in **Figure 5D**. To fix the structure together, UV light (385 nm) is shined on the whole turtle for several minutes.

*Induction heating:* The LM patterns can be inductively heated by applying external alternating magnetic fields. To achieve the plentiful heating and associated actuation performance of the LCE – LM composites, five different water-cooled coils made by hollow copper tubes are connected to a LH-15A induction heater (MXBAOHENG, China) to generate the alternating magnetic fields with different distributions and frequencies. By adjusting the currents supplied to the coils, the amplitudes of the fields can also be accurately tuned. The photos of these coils are illustrated in Figure S3 – S7. The shapes and parameters of these coils are summarized in Table S1.

**Supporting Information**

Supporting Information is available from the author.

**Acknowledgements**

The authors acknowledge the support from the National Science Foundation (NSF) Career Award CMMI-2145601 and NSF Award CNS-2201344. The authors acknowledge the useful discussions with Prof. Michael Dickey and Dr. Taylor V. Neumann on the aerosol spray deposition of liquid metal.